%% file: LHCP2014-InclusiveSUSYatATLAS.tex
\documentclass[10pt]{article}
\usepackage{graphicx}

\def\Title#1{\begin{center} {\Large #1 } \end{center}}
\def\Author#1{\begin{center}{ \sc #1} \end{center}}
\def\Address#1{\begin{center}{ \it #1} \end{center}}

\newcommand\pubblock{\rightline{\begin{tabular}{l} Proceedings of the Second Annual LHCP\\ \pubnumber\\
         \pubdate  \end{tabular}}}

\newenvironment{Abstract}{\begin{quotation} \begin{center} 
             \large ABSTRACT \end{center}\bigskip 
      \begin{center}\begin{large}}{\end{large}\end{center} \end{quotation}}

\newenvironment{Presented}{\begin{quotation} \begin{center} 
             PRESENTED AT\end{center}\bigskip 
      \begin{center}\begin{large}}{\end{large}\end{center} \end{quotation}}

\input econfmacros.tex

\usepackage{color}

\newcommand\pubnumber{ ATL-PHYS-PROC-2014-101 }

\newcommand\pubdate{\today}

\def\affiliation{
On behalf of the ATLAS Collaboration, \\
Nikhef National Institute for Subatomic Physics,\\
Amsterdam, Netherlands  \footnote{Currently at Institute of Physics, University of Belgrade, Belgrade, Serbia}\\}

\begin{document}

\large
\begin{titlepage}
\pubblock

\vfill
\Title{  Inclusive searches for squarks and gluinos with the ATLAS detector  }
\vfill

\Author{ Marija Vranje\v{s} Milosavljevi\'{c}  }
\Address{\affiliation}
\vfill
\begin{Abstract}

Despite the absence of experimental evidence, weak scale supersymmetry remains one of the best motivated and studied Standard Model extensions. This report summarises recent ATLAS results on inclusive searches for supersymmetric squarks and gluinos, including third generation squarks produced in the decay of gluinos. Results are presented for both R-parity conserving and R-parity violating scenarios, with final states containing jets with and without missing transverse momentum, light leptons, taus or photons. 

\end{Abstract}
\vfill

\begin{Presented}
The Second Annual Conference\\
 on Large Hadron Collider Physics \\
Columbia University, New York, U.S.A \\ 
June 2-7, 2014
\end{Presented}
\vfill
\end{titlepage}
\def\thefootnote{\fnsymbol{footnote}}
\setcounter{footnote}{0}
%

\normalsize

\section{Introduction}
Supersymmetry (SUSY) \cite{susyref} is one of the most promising and most studied theories for physics beyond the Standard Model (SM). 
It is a generalisation of the space-time symmetries of quantum field theory which transforms fermions into bosons and vice versa, 
while leaving the non spin-related quantum numbers unchanged. SUSY provides a natural solution to the hierarchy problem by 
cancelling out the quadratically divergent quantum corrections to the Higgs boson mass with the introduction of the 
superpartner of the top quark \cite{natural}. It also provides the unification of the three SM gauge couplings at high energy scales, 
modifying the running of the couplings above the electroweak scale. Moreover, if R-parity is conserved, the lightest supersymmetric 
particle (LSP) might be a promising candidate for a stable weakly-interacting massive particle consistent with the observed density of Dark Matter.

In these proceedings, the focus will be on the strong production of SUSY particles: scalar partners of first and
second generation quarks (squarks, $\tilde{q}$) and the fermionic partners of the gluons (gluinos, $\tilde{g}$). For a given SUSY mass scale, 
strong production has the largest production cross sections, thus allowing to target
heavy initial SUSY particles via signatures with small branching ratios and long decay chains.
Apart from one exception, all analyses presented below assume R-parity conservation, which makes the LSP stable. 
In many models this particle is the lightest neutralino $\tilde{\chi}_{0}^{1}$, or the gravitino $\tilde{G}$, and by the fact that it interacts very weakly, events are typically characterised by large missing transverse momentum, $E_{\rm T}^{\rm miss}$. 

Only results which have been made public by the ATLAS Collaboration on or before the June 2014, the date at which the corresponding talk has been given, are considered.  These are derived from analysing $\sim$ 20 $fb^{-1}$ of $pp$ collision data at $\sqrt{s}$ = 8 TeV collected by the ATLAS experiment. 
None of the analyses has seen any significant excess of events above the expected number from the SM predictions, thus they interpret their findings in terms of limits on SUSY particle masses or model parameters. Interpretations in a large number of simplified models, defined by an effective Lagrangian describing interactions of a small number of new particles, typically assuming one production process and one decay channel with a 100\% branching fraction, will be summarised, demonstrating the power of the various searches. 

\section{ Searches with many jets, missing transverse momentum and no leptons } \label {0leptonAna}
In ATLAS, these searches are split into two analyses with minimum jet multiplicities of two to six and seven to ten, respectively.

\begin{figure}[h!]
\centering
\includegraphics[width=0.33\textwidth]{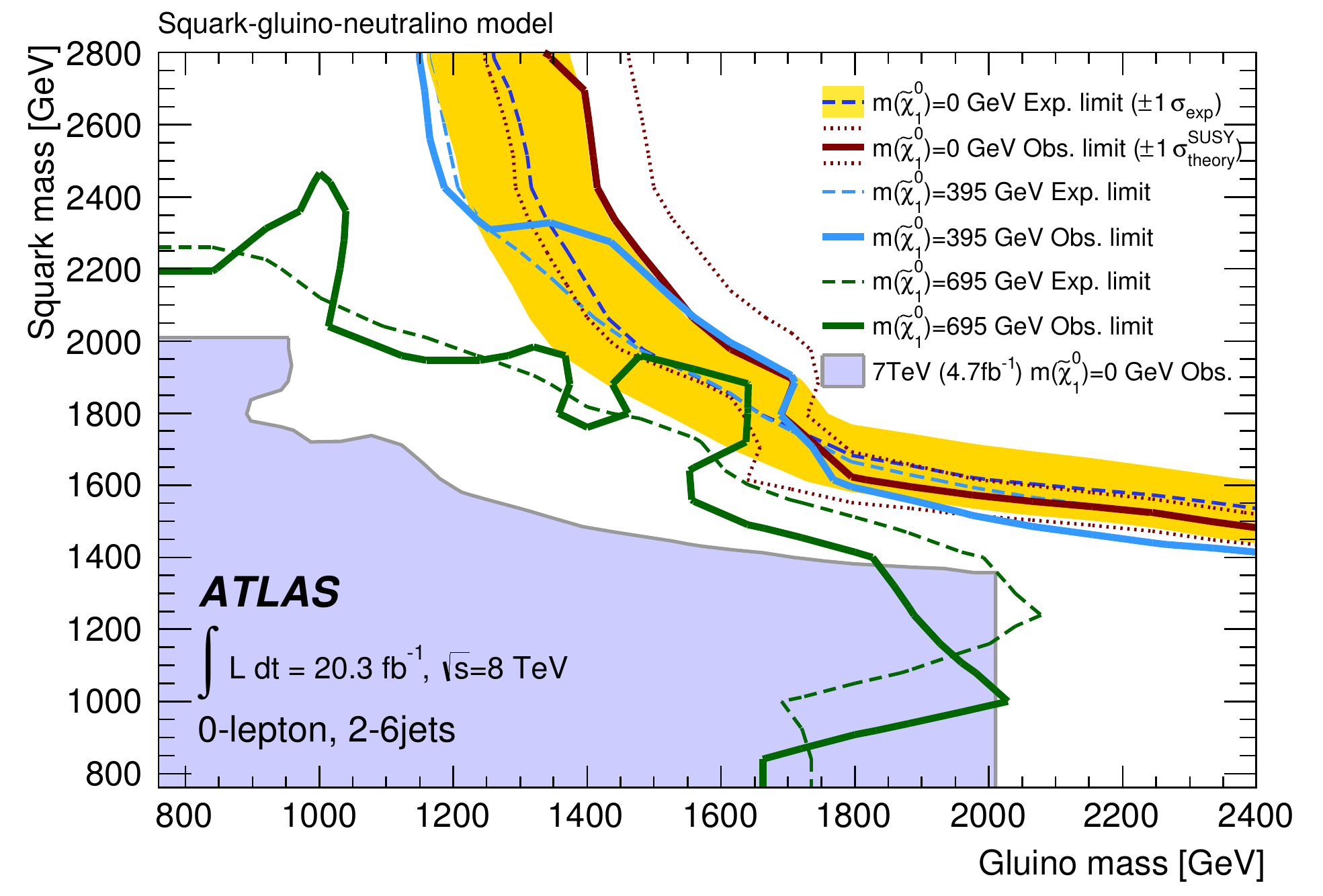}
\includegraphics[width=0.33\textwidth]{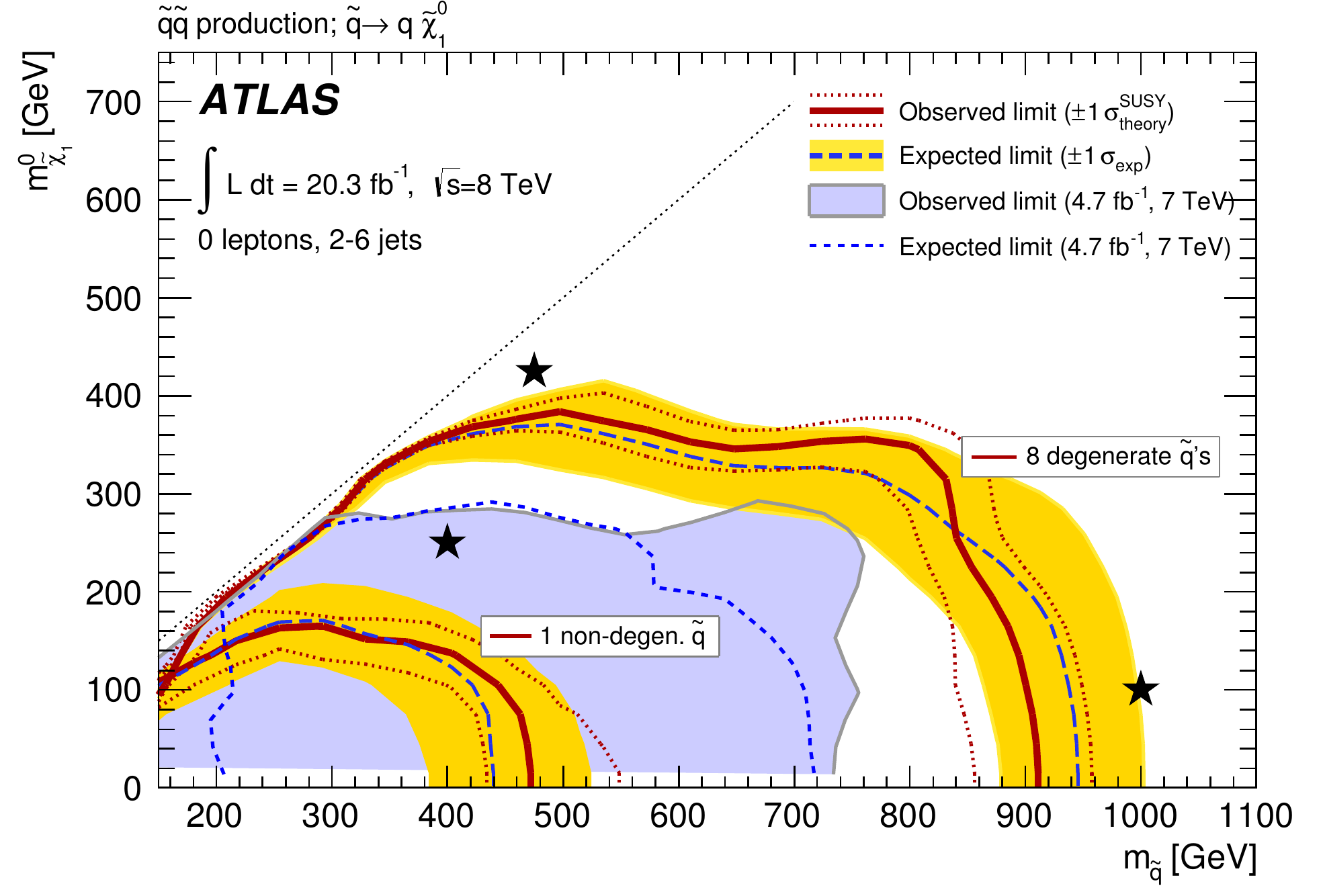}
\includegraphics[width=0.28\textwidth]{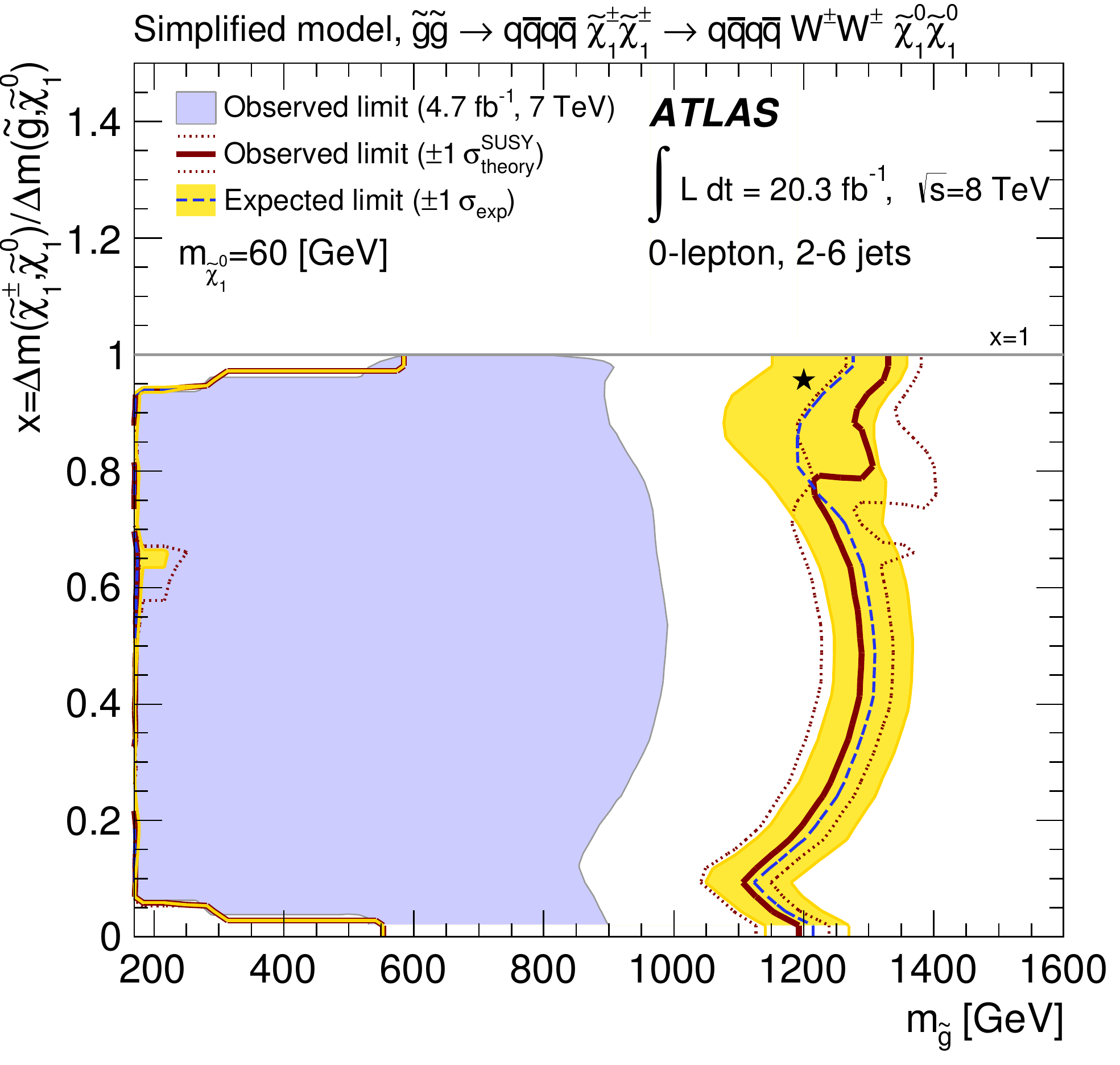}
\caption{ Observed and expected limits obtained by the analysis that requires minimum jet multiplicities of two to six, large missing transverse momentum and no leptons \cite{0lepton}, in simplified squark-gluino-neutralino MSSM model (left), simplified model involving the strong production of squarks of the first and second generations with the direct decay of the squark $\tilde{q} \rightarrow q\tilde{\chi^{0}_{1}}$ (middle) and in the models with the cascade $\tilde{g}$ decay via $\tilde{\chi}^{\pm}_{1}$ to $W$ and $\tilde{\chi}^{0}_{1}$ (right). }
\label{fig:0-lepton}
\end{figure}

The search with at least two to at least six jets \cite{0lepton} targets final states where each initial squark yields one jet plus $E_{\rm T}^{\rm miss}$, and each initial gluino yields two jets plus $E_{\rm T}^{\rm miss}$. Additional decay modes can include the production of charginos via $\tilde{q} \rightarrow q \tilde{\chi}^{\pm}_{1}$ and $\tilde{g} \rightarrow q\bar{q} \tilde{\chi}^{\pm}_{1}$, where subsequent decay of these charginos to SM $W$ boson and $\tilde{\chi}^{0}_{1}$ can lead to final states with larger jet multiplicities. In this analysis, events with reconstructed electrons or muons are vetoed and the search strategy is optimised in gluino-squark mass plane for a range of models. 
Fifteen inclusive signal regions are characterised by increasing minimum jet-multiplicity from two to at least six, and are based on different 
selection cuts on the effective mass ${m_{\rm eff}}$, defined as the scalar sum of  $E_{\rm T}^{\rm miss}$ and the $p_{\rm T}$ of the jets,  
the ratio of $E_{\rm T}^{\rm miss}$/${m_{\rm eff}}$, and the minimum azimuthal angle between jets and $E_{\rm T}^{\rm miss}$ in the transverse plane. 
Two of the signal regions are designed to improve sensitivity to models with the cascade $\tilde{q}$ or $\tilde{g}$ decay via $\tilde{\chi}^{\pm}_{1}$ to $W$ and $\tilde{\chi}^{0}_{1}$, in cases where the $\tilde{\chi}^{\pm}_{1}$ is nearly degenerate in mass with the $\tilde{q}$ or $\tilde{g}$. 
These signal regions place additional requirements on the invariant masses $m(W_{\rm cand})$ of the candidate $W$ bosons reconstructed from single high-mass jets, or from the pairs of jets. 
The limits on $\tilde{q}$ and $\tilde{g}$ masses are found to be similar for the mSUGRA/CMSSM implementation and
a simplified squark-gluino-neutralino MSSM model presented in Figure \ref{fig:0-lepton} (left), excluding equal mass light-flavor squarks and gluinos with masses below 1.65 TeV. Results interpreted within a simplified model involving the strong production of squarks of the first and second generations and in the models with the cascade $\tilde{q}$ or $\tilde{g}$ decay via $\tilde{\chi}^{\pm}_{1}$ to $W$ and $\tilde{\chi}^{0}_{1}$ are also shown in Figure \ref{fig:0-lepton}.   

\begin{figure}[h!]
\centering
\includegraphics[width=0.31\textwidth]{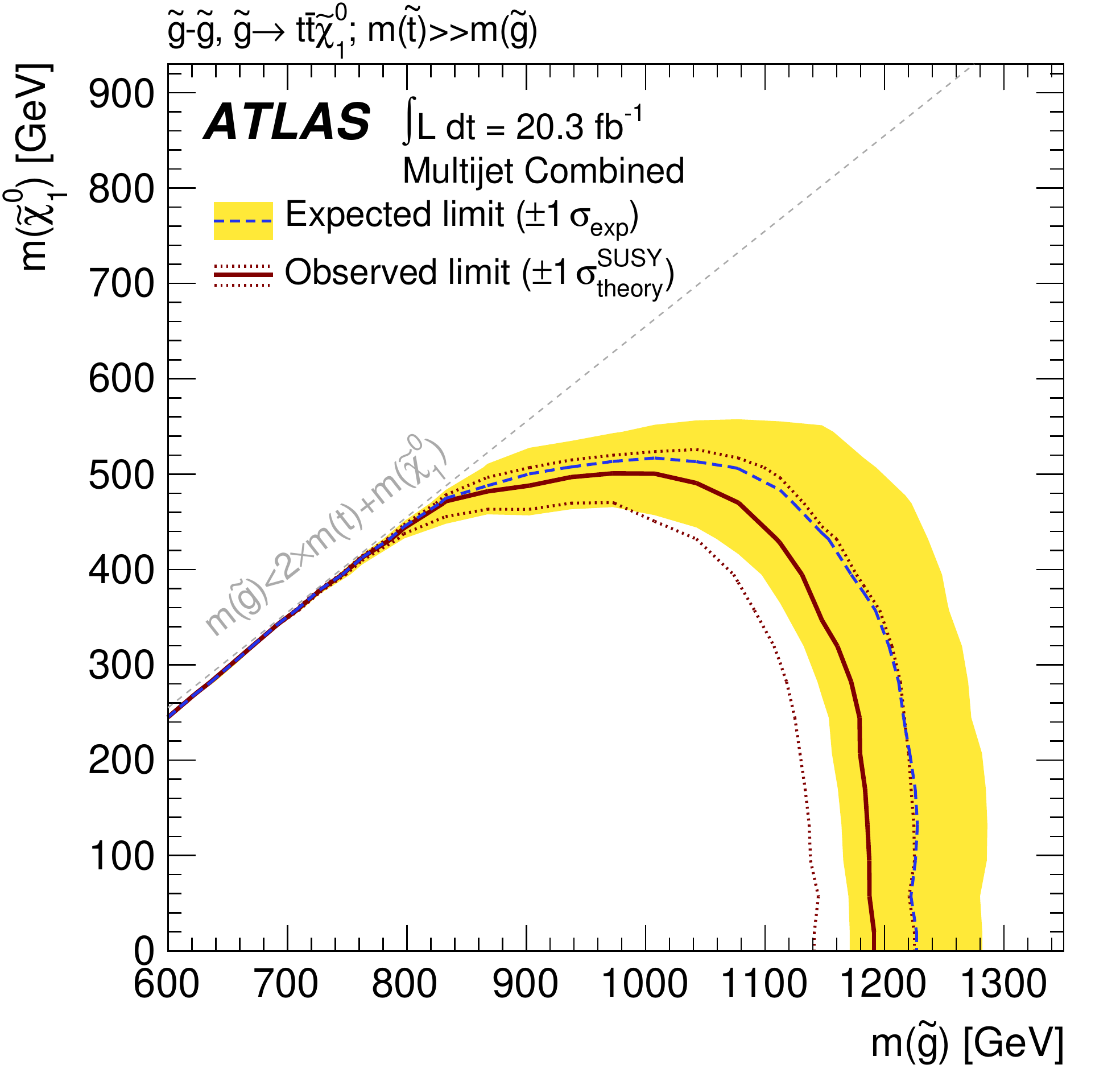}
\includegraphics[width=0.31\textwidth]{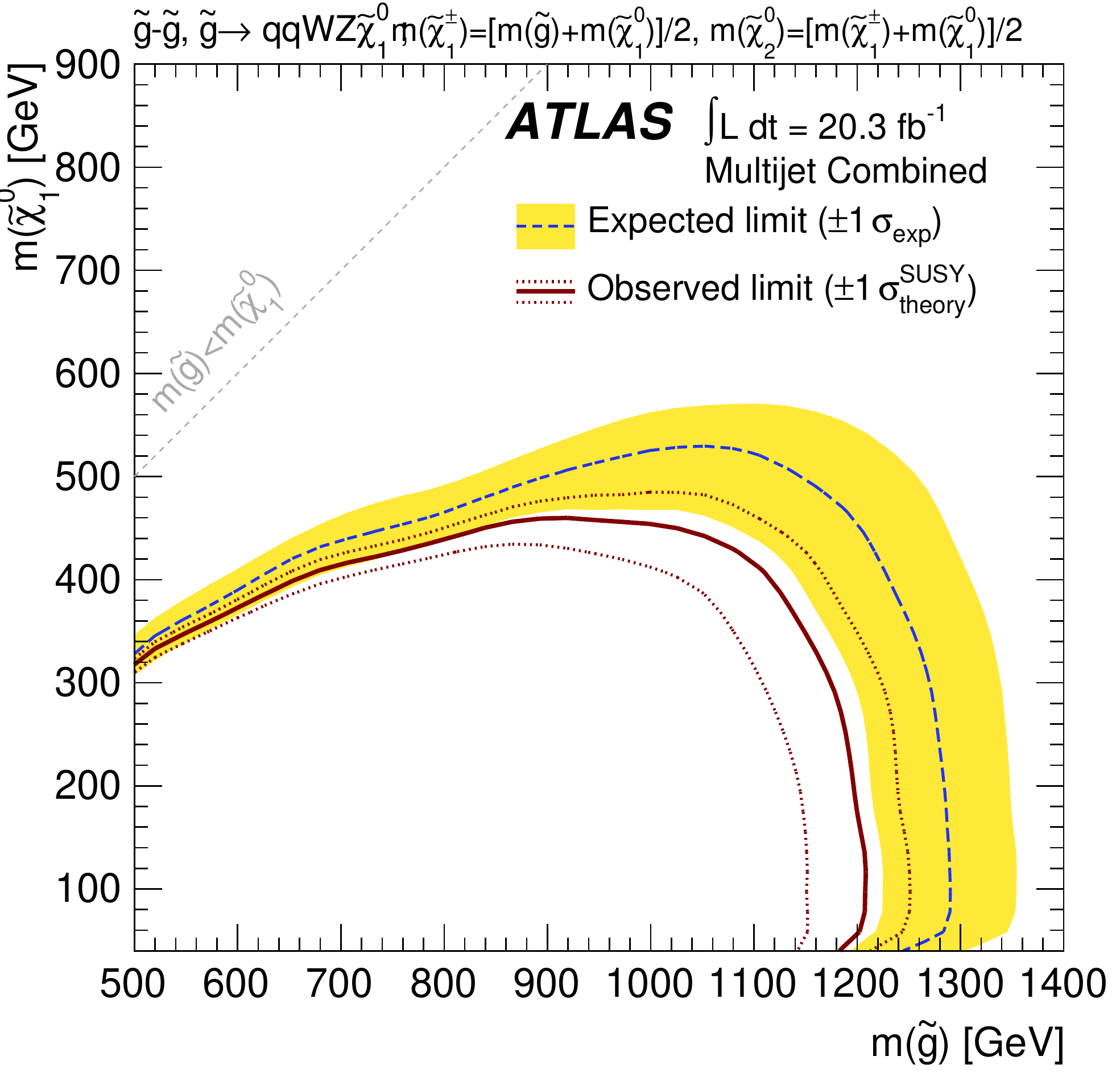}
\caption{ Observed and expected limits obtained by the analysis with minimum jet multiplicities of seven to at least ten, large missing transverse momentum and no leptons \cite{multijets}, in simplified model which contains only gluino octet and a neutralino assuming gluino decay $\tilde{g} \rightarrow t \bar{t} \tilde{\chi}^{0}_{1}$ via an off-shell $\tilde{t}$-squark (left), and in simplified model where each gluino decays as $\tilde{g} \rightarrow \bar{q}+q'+\tilde{\chi}^{\pm} \rightarrow \bar{q}+q'+W^{\pm}+\tilde{\chi}^{0}_{2} \rightarrow \bar{q}+q'+W^{\pm}+Z^{0}+\tilde{\chi}^{0}_{1}$ (right). }
\label{fig:multijets}
\end{figure}

High jet multiplicities are expected from the decays of gluino pairs via top squark, or via squarks involving the production of $\tilde{\chi}^{\pm}$ or $\tilde{\chi}^{\pm}$ and $\tilde{\chi}^{0}_{2}$, and is the main topology targeted by the analysis with at least seven to at least ten jets, significant $E_{\rm T}^{\rm miss}$ and the absence of isolated electrons or muons \cite{multijets}. 
The sensitivity of the search is enhanced by the subdivision into several categories. Firstly, event classification based on the number of $b$-jets gives enhanced sensitivity to models which predict either more or fewer $b$-jets than the SM background.
Secondly, in a complementary stream of the analysis, the jets reconstructed by the jet radius parameter R = 0.4 are reclustered into large composite jets (R = 1.0) to form an event variable, the sum of the masses of the composite jets $M_{J}^{\Sigma}$, which gives additional discrimination in models with a large number of objects in the final state.  
The fully data-driven background determination method is based on the observation that the $E_{\rm T}^{\rm miss}$ resolution is approximately proportional to $\sqrt{H_{\rm T}}$ (where $\sqrt{H_{\rm T}}$ is the scalar sum of $p_{\rm T}$ of all jets), and almost independent of the jet multiplicity in events dominated by the jet activity.  
The analysis result interpreted in a simplified model that contains only gluino octet and a neutralino within the kinematic reach, assuming gluino decay $\tilde{g} \rightarrow t \bar{t} \tilde{\chi}^{0}_{1}$ via an off-shell $\tilde{t}$-squark, is shown in Figure \ref{fig:multijets} (left).
Results within a simplified model where each gluino decays promptly via an off-shell squark as $\tilde{g} \rightarrow \bar{q}+q'+\tilde{\chi}^{\pm} \rightarrow \bar{q}+q'+W^{\pm}+\tilde{\chi}^{0}_{2} \rightarrow \bar{q}+q'+W^{\pm}+Z^{0}+\tilde{\chi}^{0}_{1}$, are shown in Figure \ref{fig:multijets} (right).

\section{ Searches with leptons, jets and missing transverse momentum } \label {1leptonAna}
Two analyses addressing inclusive searches for squarks and gluinos in events with leptons, jets and missing transverse momentum will be summarised here: searches with an isolated lepton, and searches in final states with the same-sign lepton pairs or at least three leptons. 

\begin{figure}[h!]
\centering
\includegraphics[width=0.38\textwidth]{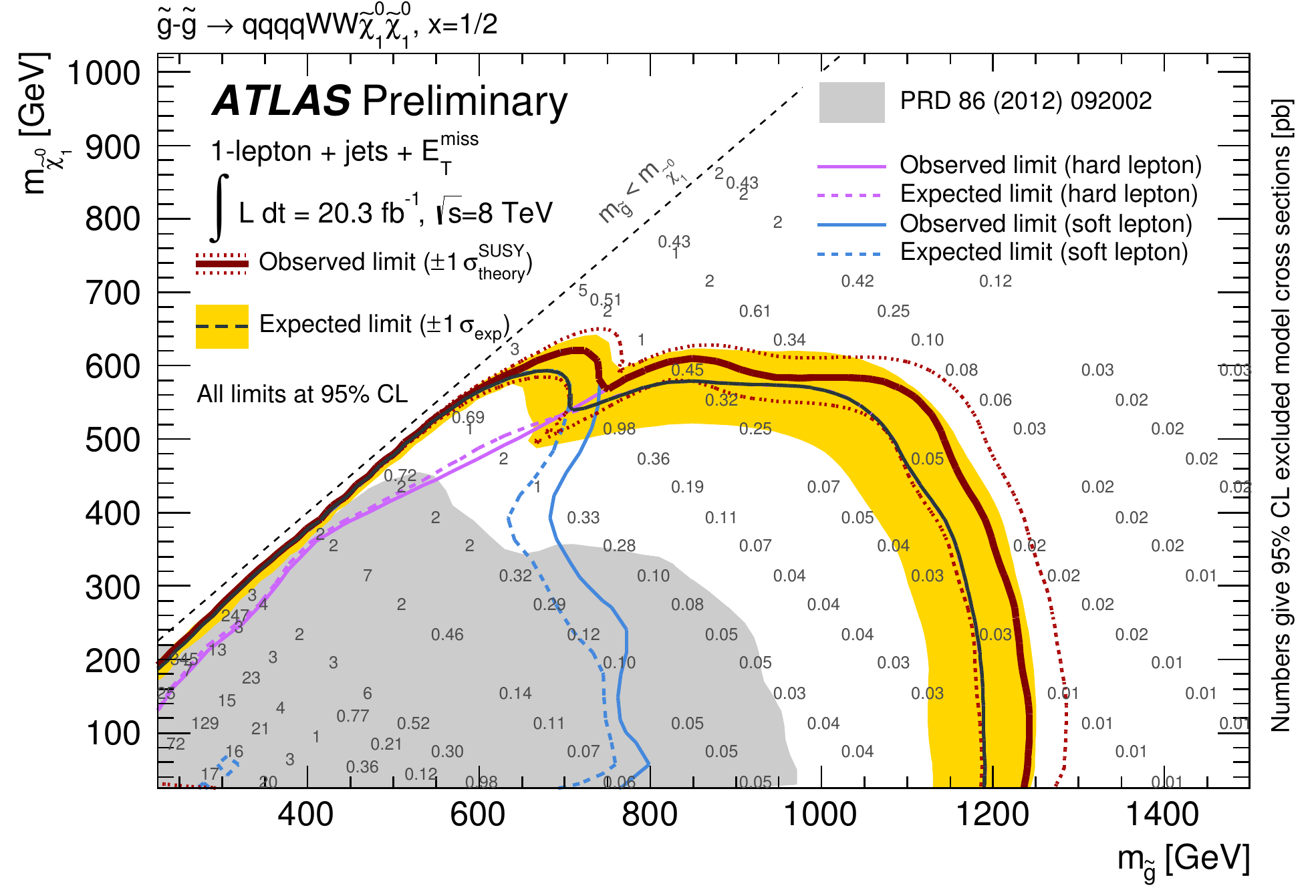}
\includegraphics[width=0.38\textwidth]{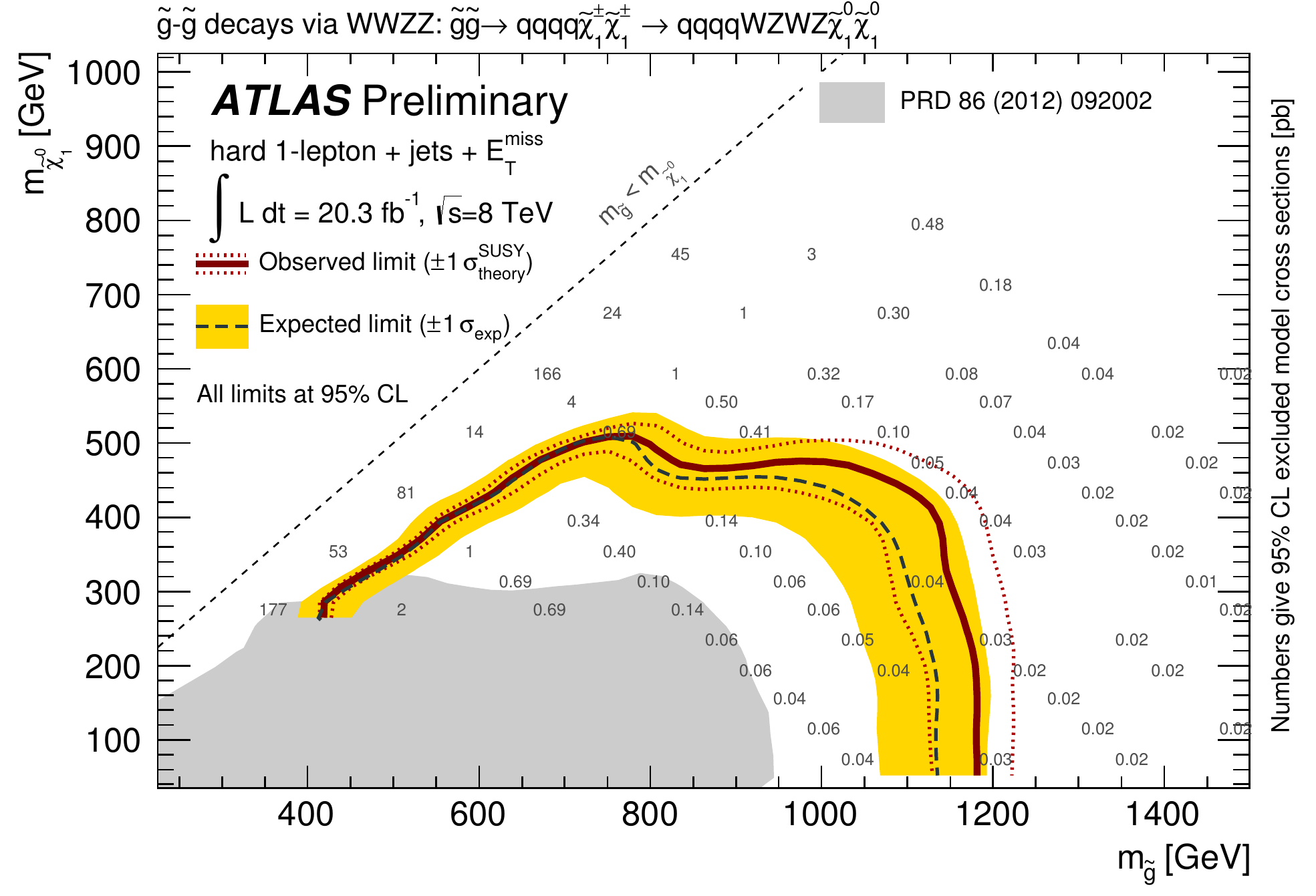}
\caption{ Observed and expected limits obtained by the analysis with an isolated lepton, jets and missing transverse momentum \cite{1-lepton}, in simplified model with the cascade $\tilde{g}$ decay via $\tilde{\chi}^{\pm}_{1}$ to $W$ and $\tilde{\chi}^{0}_{1}$ (left), and in case $\tilde{g}$ decays as $\tilde{g} \rightarrow \bar{q}+q'+\tilde{\chi}^{\pm} \rightarrow \bar{q}+q'+W^{\pm}+\tilde{\chi}^{0}_{2} \rightarrow \bar{q}+q'+W^{\pm}+Z^{0}+\tilde{\chi}^{0}_{1}$ (right). }
\label{fig:1lepton}
\end{figure}

First analysis \cite{1-lepton} uses a statistically independent set of events, compared to analyses previously described in Section \ref{0leptonAna}, by requesting the presence of one isolated lepton (electron or muon) in the event. This analysis uses events characterised by the presence of low (10(6) $< p_{\rm T} \leq$ 25 GeV) or high ($p_{\rm T} >$ 25 GeV) transverse momentum electrons (muons), typically referred to as the soft-lepton and the hard-lepton channels. 
Soft-lepton signal regions are optimised to be sensitive to the models with compressed spectra of SUSY particles, where the lightest chargino is nearly mass-degenerate with the lightest neutralino (Figure \ref{fig:1lepton} (left)). Several signal regions are defined in each channel, based on jet multiplicity and different selection cuts on $E_{\rm T}^{\rm miss}$, ${m_{\rm eff}}$ and $m_{\rm T}$, where $m_{\rm T}$ is the transverse mass of the lepton and the $E_{\rm T}^{\rm miss}$. 
Exclusion limits within simplified models involving leptons in final states are given in Figure \ref{fig:1lepton} and are comparable to results obtained by the two analyses described in Section \ref{0leptonAna}. 

\begin{figure}[h]
\centering
\includegraphics[width=0.27\textwidth]{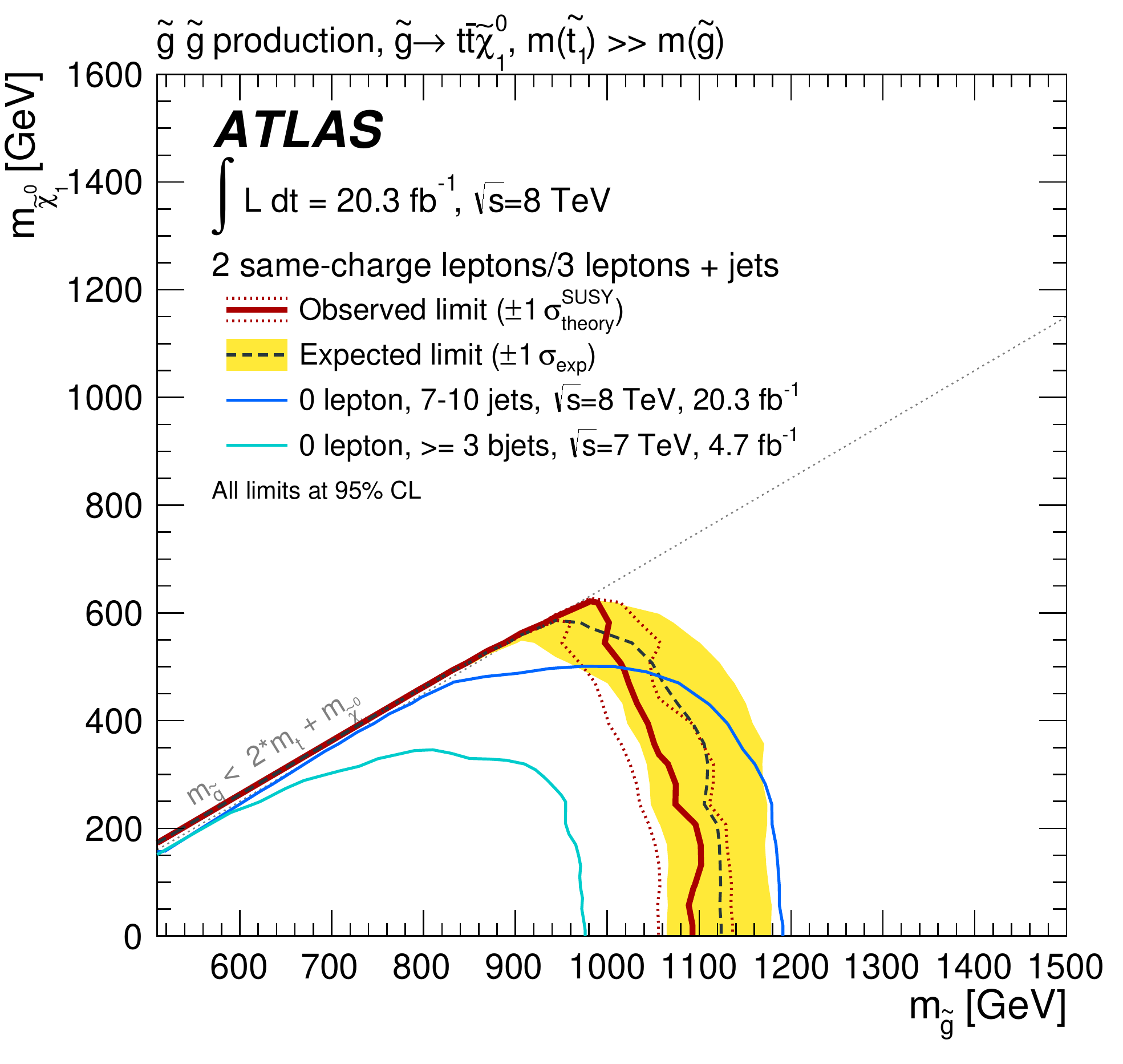}
\includegraphics[width=0.27\textwidth]{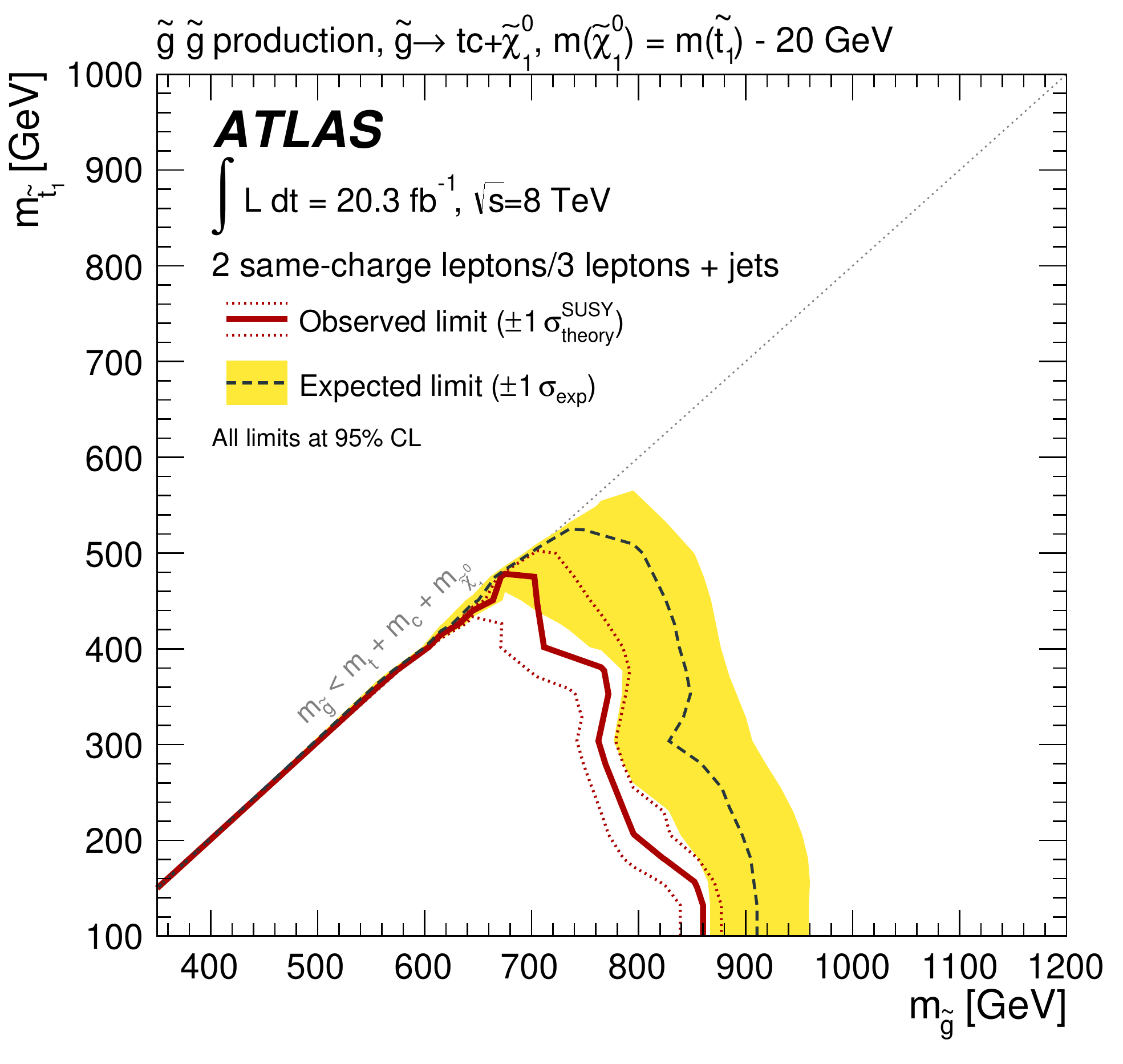}
\caption{ Observed and expected limits obtained by the analysis with same-sign or tree leptons and multiple jets \cite{same-sign}, in simplified model where gluino mediated top squark decays as $\tilde{t_{1}} \rightarrow t \tilde{\chi}^{0}_{1}$  (left), or $\tilde{t_{1}} \rightarrow c \tilde{\chi}^{0}_{1}$  (right). }
\label{fig:same-sign-gtt}
\end{figure}

The second analysis \cite{same-sign} looks at the final states with multiple jets, and either two leptons of the same electric charge (same-sign leptons) or at least three leptons to search for the strongly produced supersymmetric particles.
The motivation for this analysis is that the pair-produced gluinos have almost the same probability to give pairs of leptons with same charge or with opposite charge.
Requiring a pair of leptons with same sign suppresses the background coming from SM processes considerably, giving a very clean and powerful signature to look for new physics processes. It also allows the use of relatively loose kinematic requirements on $E_{\rm T}^{\rm miss}$, increasing the sensitivity to scenarios with small mass differences between SUSY particles.
Results for simplified models where gluinos are produced in pairs and the top squark is assumed to be the lightest squark are presented in Figure \ref{fig:same-sign-gtt}, for the case where $\tilde{t}$ decays as $\tilde{t_{1}} \rightarrow t \tilde{\chi}^{0}_{1}$ (left) or as $\tilde{t_{1}} \rightarrow c \tilde{\chi}^{0}_{1}$ (right). 

\section{ Searches with $b$-jets and missing transverse momentum }
\begin{figure}[h]
\centering
\includegraphics[width=0.32\textwidth]{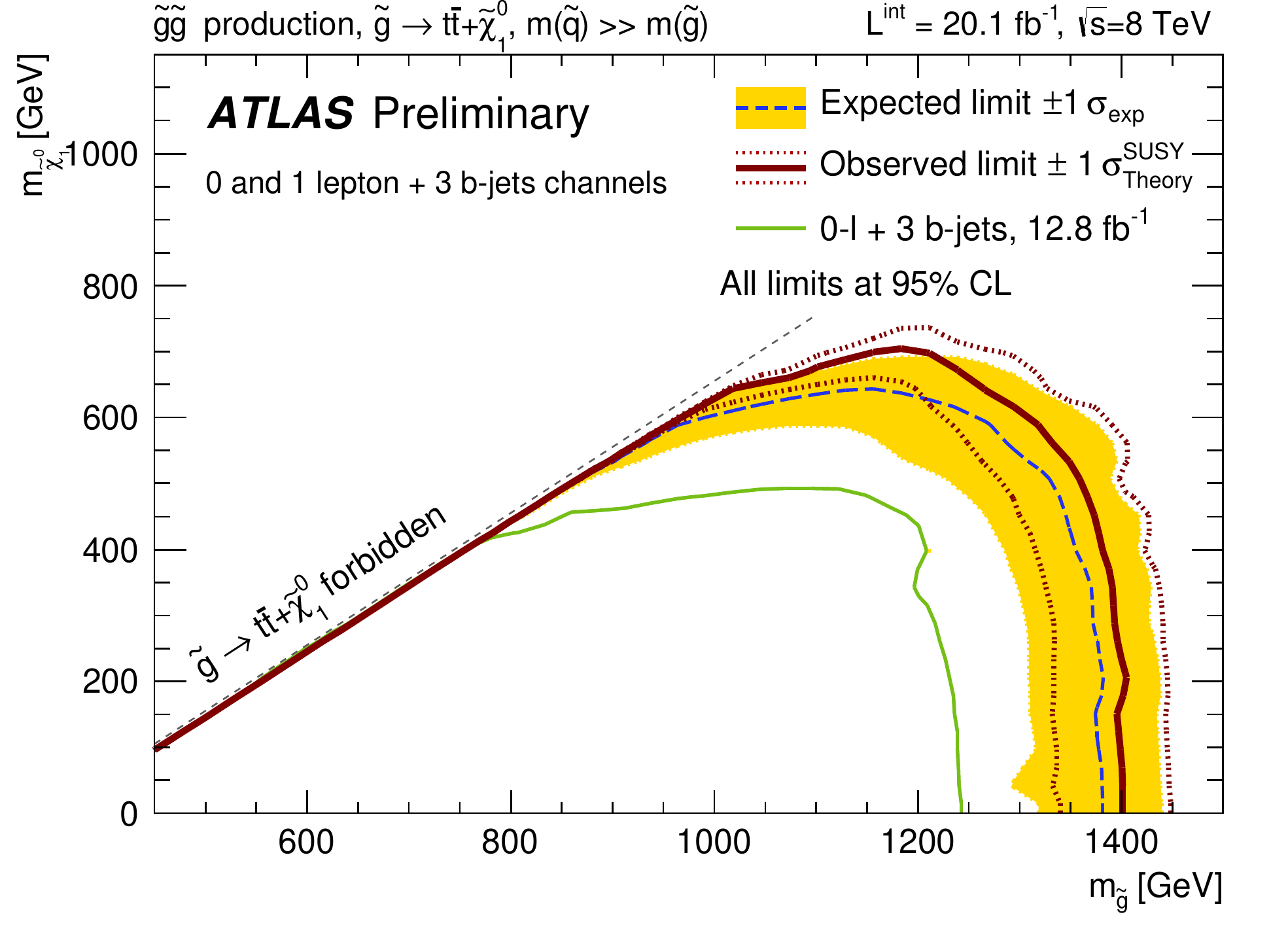}
\includegraphics[width=0.32\textwidth]{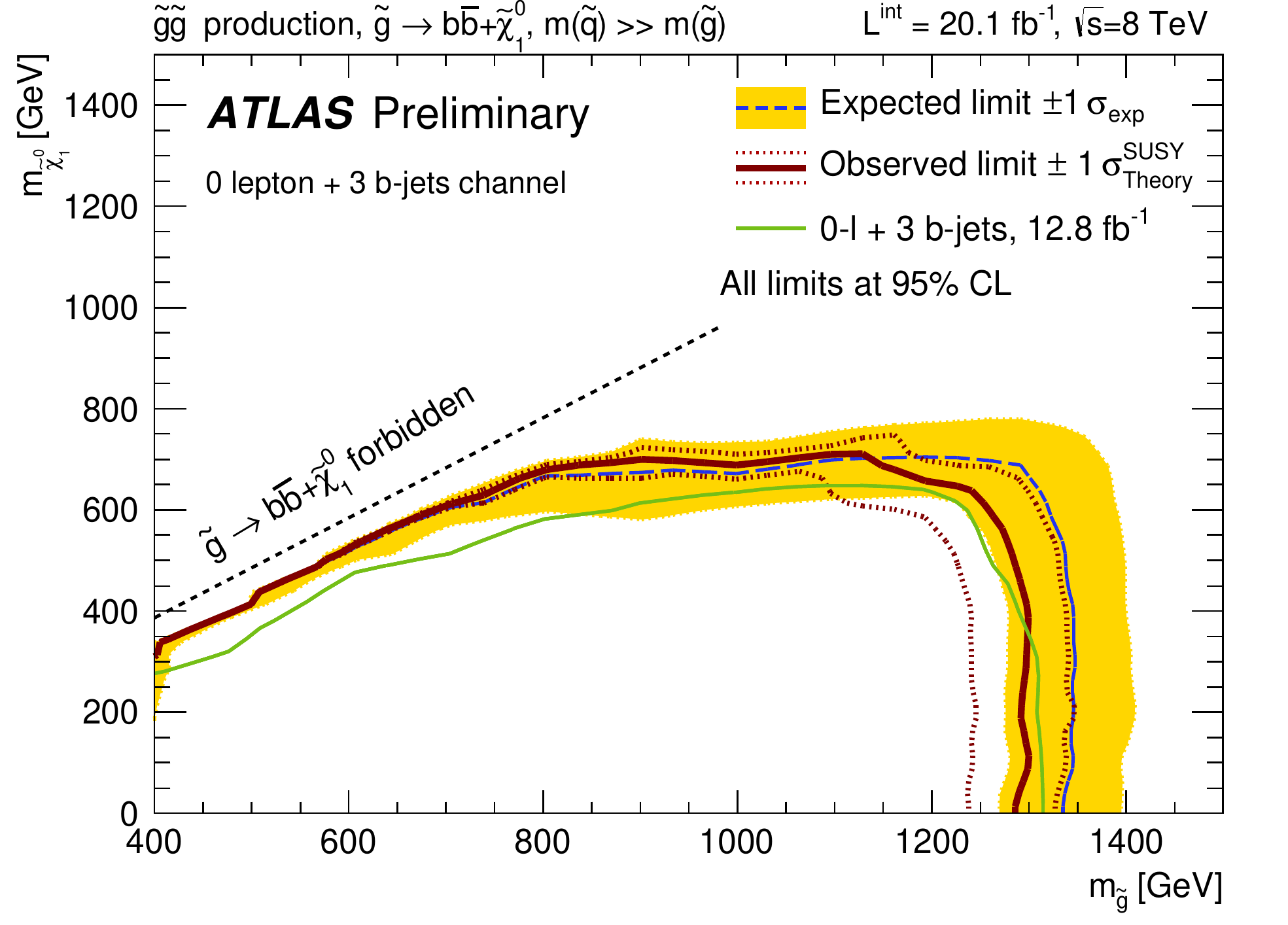}
\includegraphics[width=0.32\textwidth]{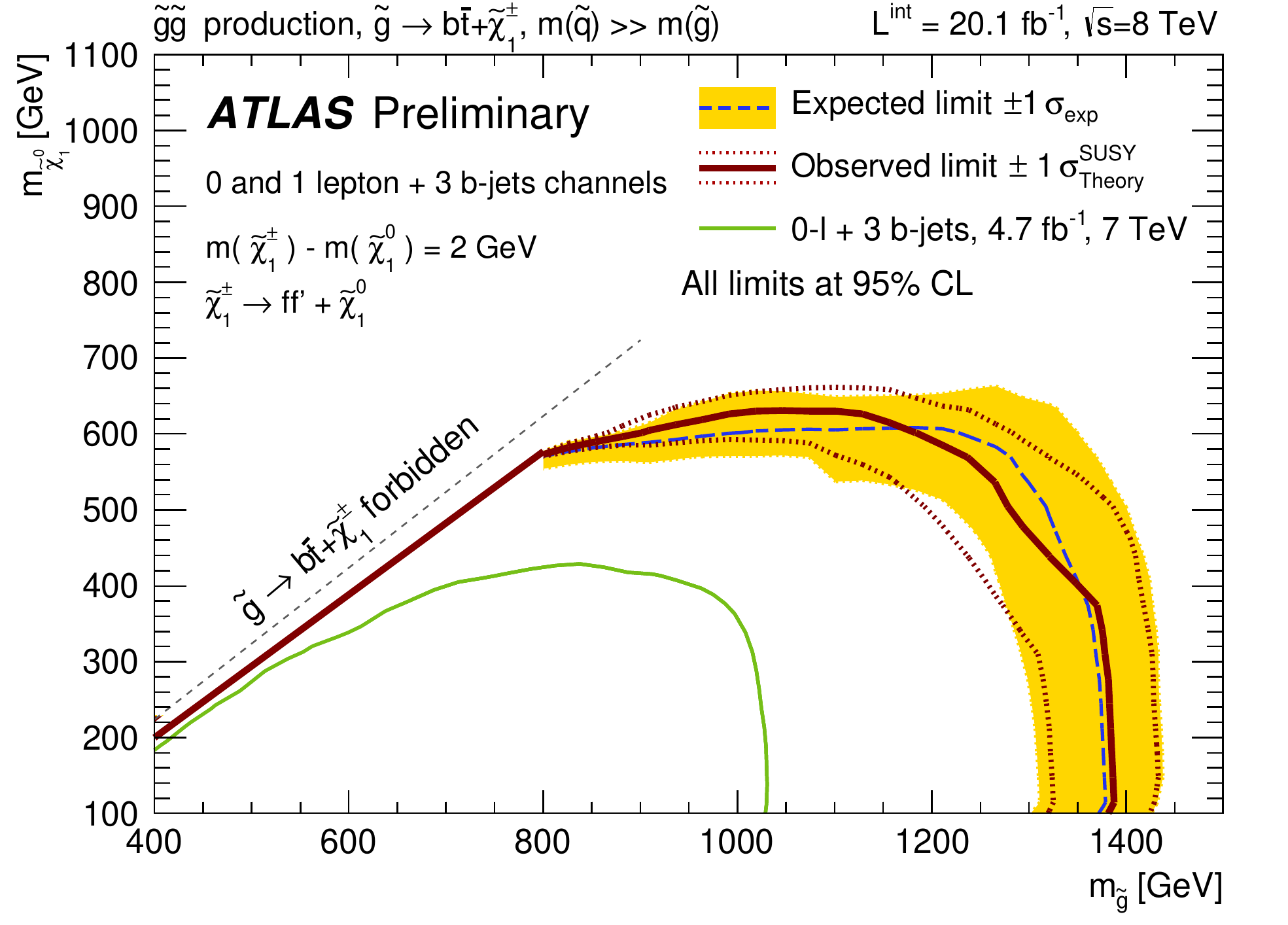}
\caption{ Observed and expected limits obtained by the analysis that requires presence of at least tree b-jets in event \cite{3b}, in simplified model where gluinos are produced in pairs and decay as $\tilde{g} \rightarrow t\bar{t}+\tilde{\chi}^{0}_{1}$ (left),  $\tilde{g} \rightarrow b\bar{b}+\tilde{\chi}^{0}_{1}$ (middle) or $\tilde{g} \rightarrow b\bar{t}+\tilde{\chi}^{0}_{1}$ (right).}
\label{fig:3b}
\end{figure}

The search that requires at least three jets tagged as originating from $b$-quarks \cite{3b}, is one of the most sensitive searches to various SUSY models where top or bottom quarks are produced in the gluino decay chains. The analysis is carried out in the orthogonal zero- and one-lepton channels which are combined to maximise the sensitivity. All reducible background sources originating from events with at least one mis-tagged $b$-jet are estimated simultaneously using a data-driven matrix method which consists of solving a system of equations based on the number of $b$-tagged and non $b$-tagged jets in each event, along with the b-tagging efficiency and mis-tag rate. Exclusion limits in simplified models where pair-produced gluino decays as $\tilde{g} \rightarrow t\bar{t}+\tilde{\chi}^{0}_{1}$,  $\tilde{g} \rightarrow b\bar{b}+\tilde{\chi}^{0}_{1}$ or $\tilde{g} \rightarrow b\bar{t}+\tilde{\chi}^{0}_{1}$ are given in Figure \ref{fig:3b}.

\section{ Searches with taus or photons }
\begin{figure}[h!]
\centering
\includegraphics[width=0.3\textwidth]{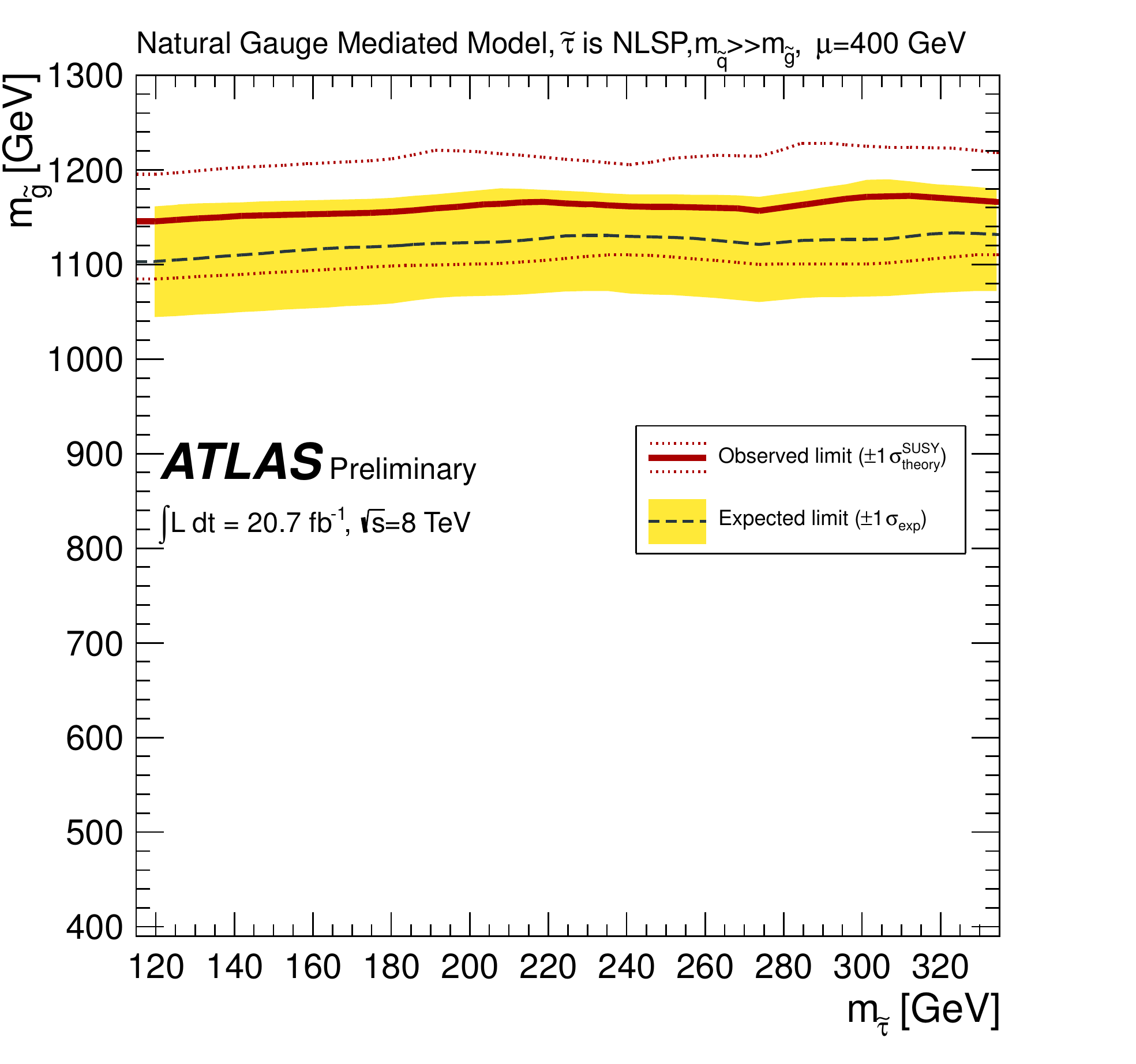}
\includegraphics[width=0.44\textwidth]{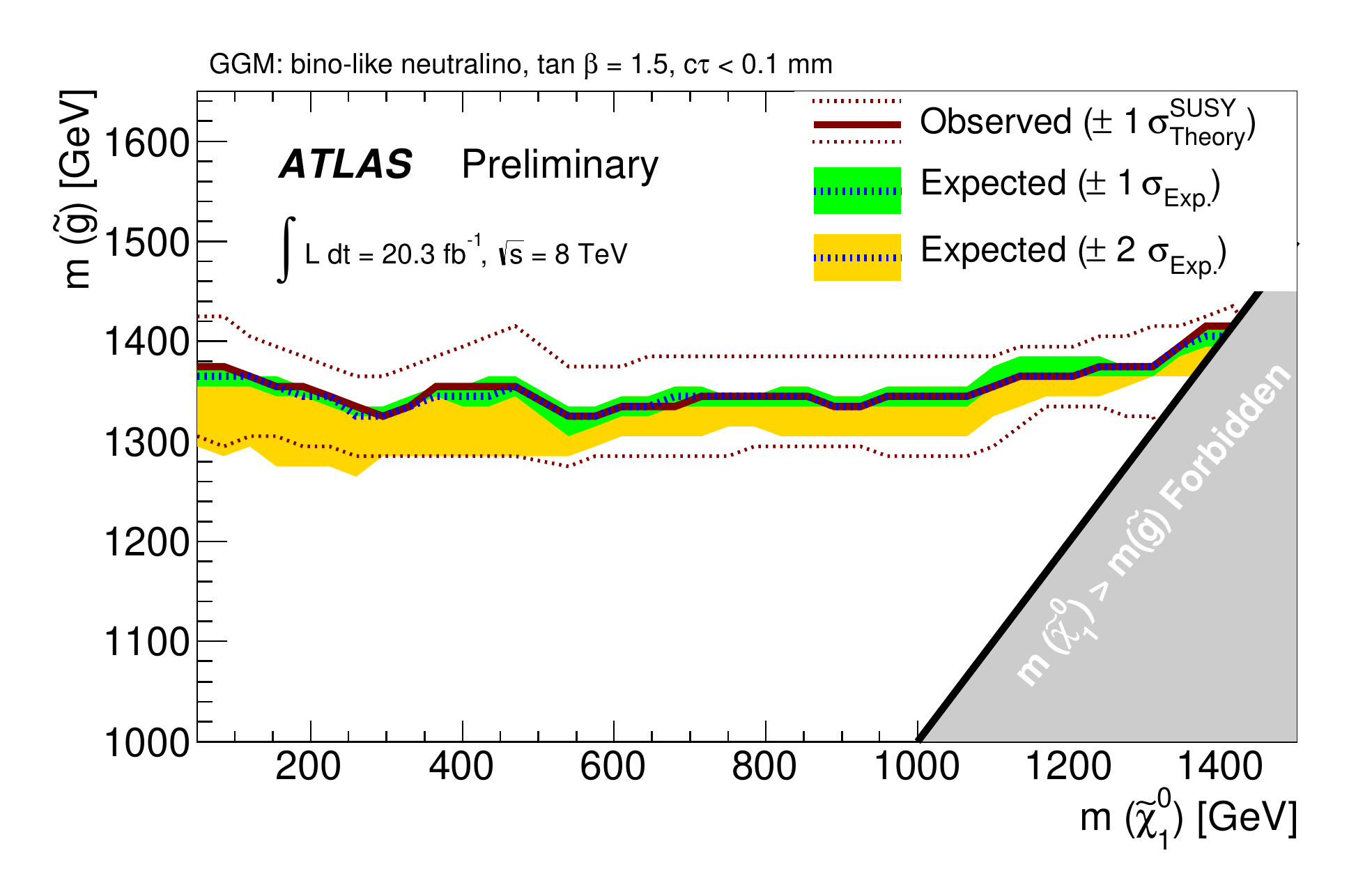}
\caption{ Observed and expected limits obtained by the analyses that require presence of at least one tau lepton \cite{tauAna} (left) or two photons \cite{Diphoton} in event (right) within the framework of general gauge mediation (GGM). }
\label{fig:TauPhoton}
\end{figure}

These searches belong to the searches for gauge-mediated models of supersymmetry breaking, where the breaking is mediated from the invisible sector (where breaking takes place) to the electroweak scale by gauge interactions. The lightest SUSY particle is gravitino $\tilde{G}$, and a number of candidates can be the next-to-lightest SUSY particle (NLSP). Analysis described in \cite{tauAna} assumes that the NLSP is stau $\tilde{\tau}$, while the analysis described in \cite{Diphoton} assumes that the NLSP is neutralino $\tilde{\chi}_{0}^{1}$, leading, respectively, to tau plus $E_{\rm T}^{\rm miss}$, or diphoton plus $E_{\rm T}^{\rm miss}$ final states. 
Results of these two analyses within the framework of general gauge mediation (GGM) are given in Figure \ref{fig:TauPhoton}.

\section{ Searches for RPV models based on signatures with many jets }
\begin{figure}[h!]
\centering
\includegraphics[width=0.27\textwidth]{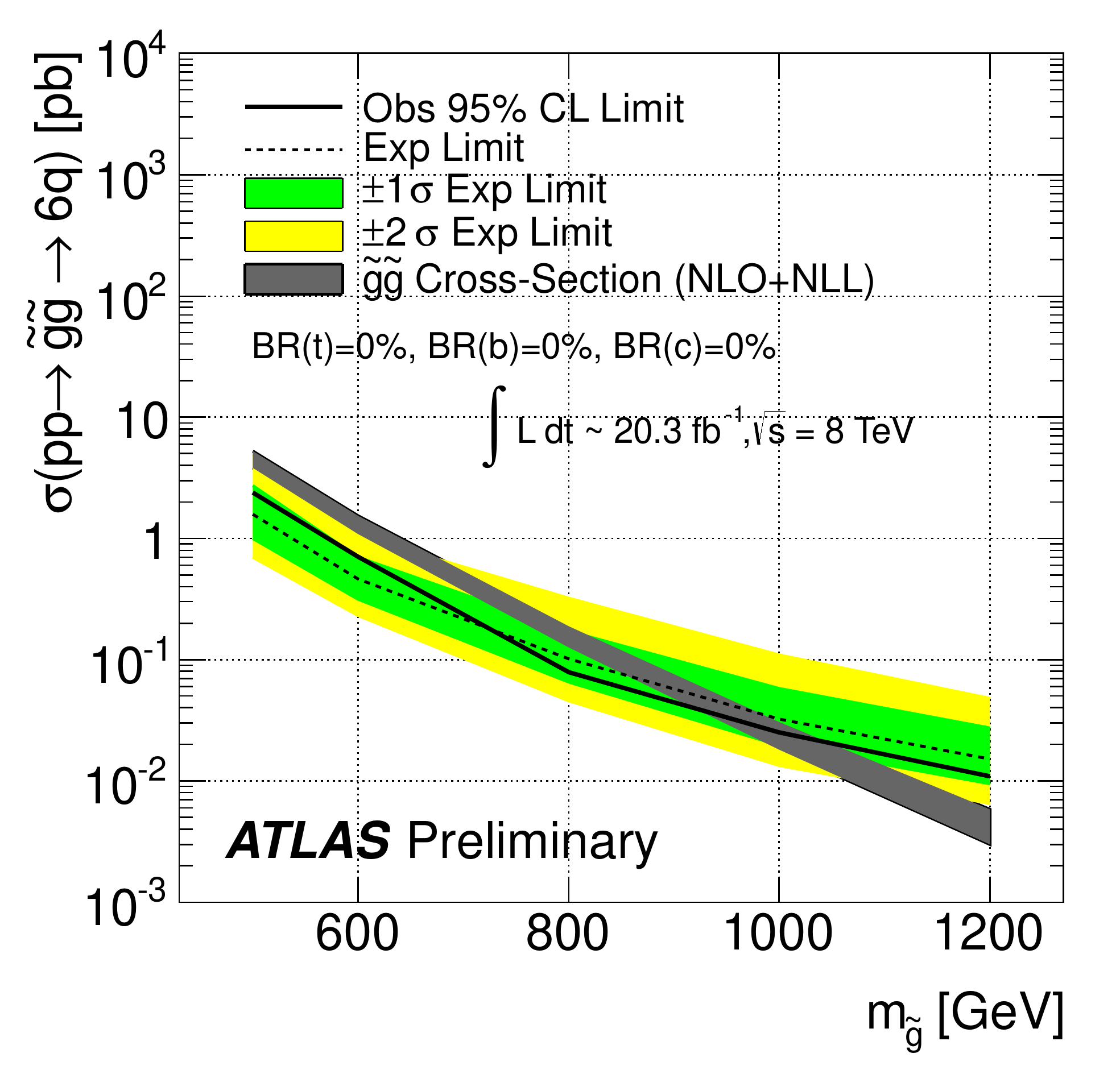}
\includegraphics[width=0.27\textwidth]{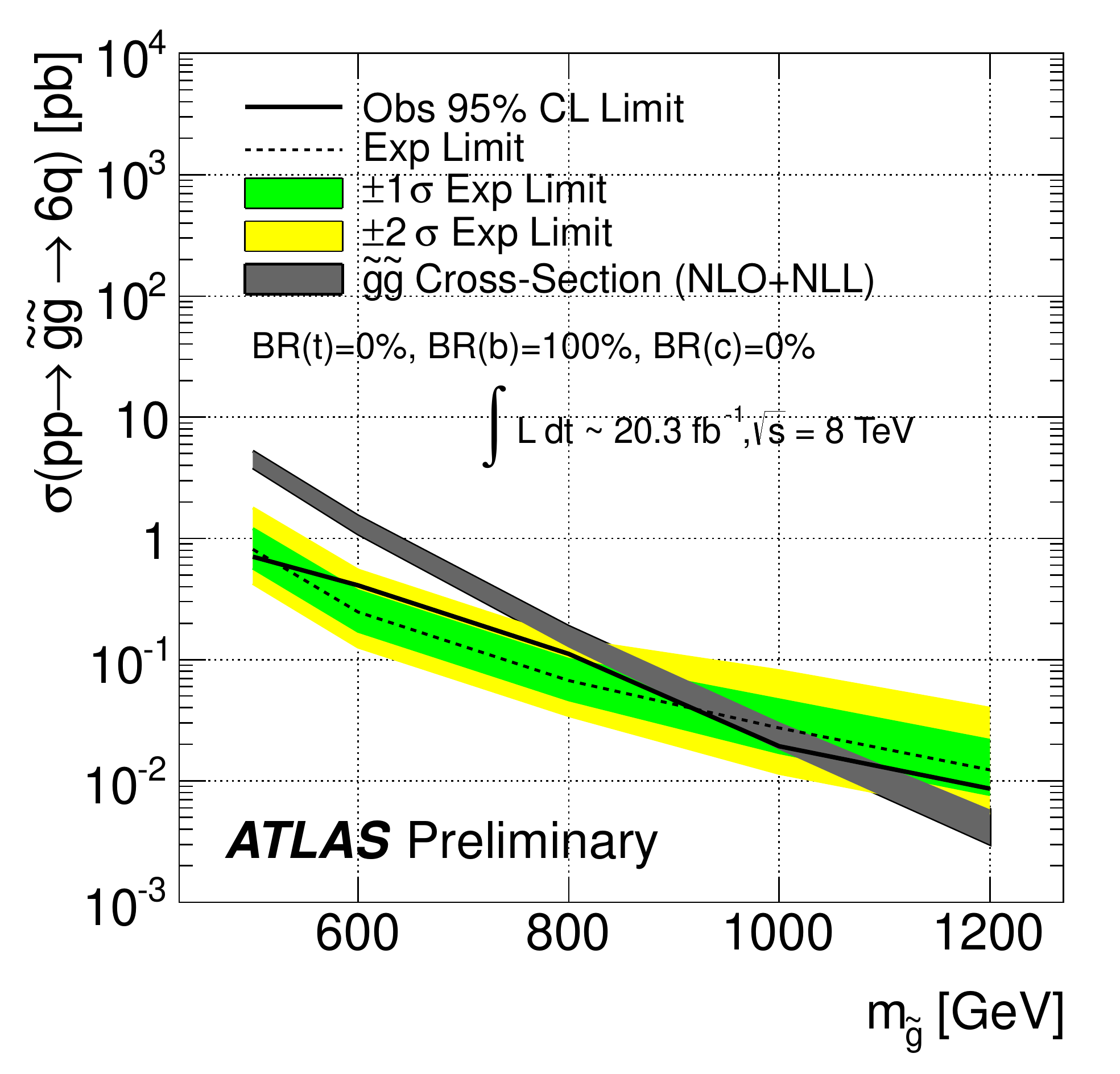}
\includegraphics[width=0.27\textwidth]{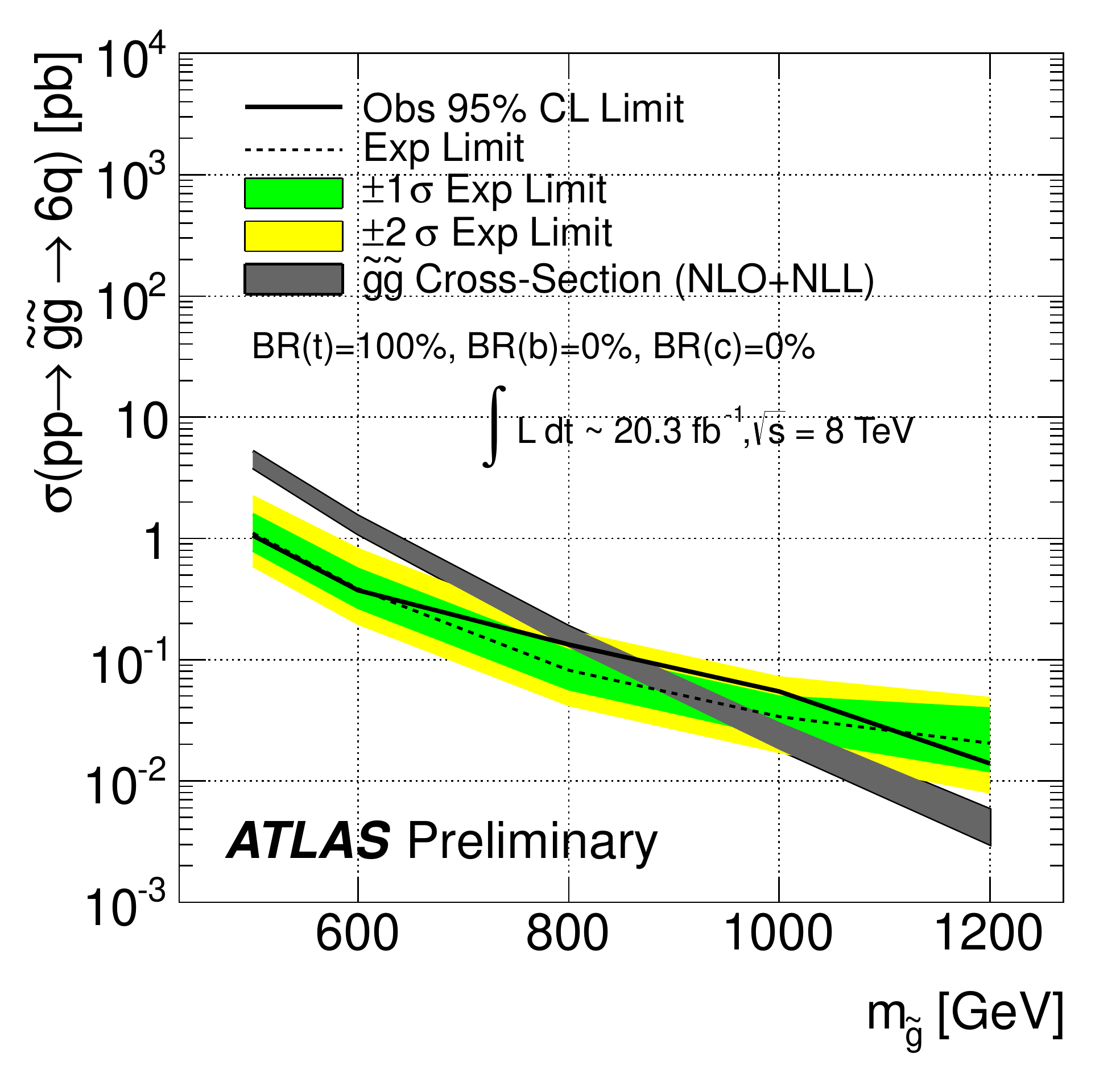}
\caption{ Expected and observed cross-section limits for the RPV 6-quark gluino models \cite{RPVmultijets}, for the case where no gluinos decay into heavy-flavour quarks (left), where gluinos decay into a b-quark (middle), or into a top quark in the final state (right). }
\label{fig:rpv}
\end{figure}

Searches summarised in \cite{RPVmultijets} are designed to look for a new physics in various final states with large number of high $p_{\rm T}$ jets, and provide results interpreted in the context of R-parity violating (RPV) supersymmetric models: the "6-quark model" in which gluinos are pair-produced and decay promptly via a virtual squark in the cascade decay $\tilde{g} \rightarrow \tilde{q} q \rightarrow qqq$, (presented here in Figure \ref{fig:rpv}), and the "10-quark model", in which gluinos decay via an intermediate on-shell neutralino in the process $\tilde{g} \rightarrow \tilde{q} q \rightarrow q q \tilde{\chi}_{0}^{1} \rightarrow qqqqq$. Results are presented for all possible RPV branching fractions of gluino decays to various quark flavours. 

\section{Conclusions}
The ATLAS Collaboration has conducted many searches for the production of squarks and gluinos using $\sim$ 20 $fb^{-1}$ of $pp$ collision data at $\sqrt{s}$ = 8 TeV in events with final states containing jets, light leptons, taus or photons, with and without missing transverse momentum. 
None of the analyses has observed a significant deviation from SM predictions, leading to constraints on many SUSY models. 
After the 2013-2014 LHC shutdown, the LHC center-of-mass energy is expected to be increased to at least 13 TeV. Given the expected large integrated luminosity which will be collected, a whole new era of searches for the supersymmetric particles will start in 2015.



\end{document}

%% file: econfmacros.tex
\def\beq{\begin{equation}}
\def\eeq#1{\label{#1}\end{equation}}
\def\eeqn{\end{equation}}

\def\beqa{\begin{eqnarray}}
\def\eeqa#1{\label{#1}\end{eqnarray}}
\def\eeqan{\end{eqnarray}}

\let\bar=\overbar

\def\Dslash{\not{\hbox{\kern-4pt $D$}}}
\def\dslash{\not{\hbox{\kern-2pt $\del$}}}

\def\msb{{\bar{\ssstyle M \kern -1pt S}}}

